\documentclass[12pt]{article}

\usepackage{natbib}
\usepackage[nottoc,notlof,notlot]{tocbibind} 

\bibpunct{(}{)}{;}{a}{}{;}
\usepackage{breakcites}
\usepackage{graphicx,xcolor}
\graphicspath{ {images/} }
\usepackage[algoruled]{algorithm2e}
\usepackage[pagebackref=false,breaklinks=true]{hyperref} 
\hypersetup{
    colorlinks=true,
    citecolor=black,
    filecolor=black,
    linkcolor=black,
    urlcolor=black,
    bookmarksopen=true,
    pdfstartview=FitH
}
\usepackage{caption,threeparttable}

\usepackage{amsthm,amsmath}
\usepackage{bbm}
\usepackage{amsfonts}%
\usepackage{amssymb}%

\usepackage{geometry} 
\usepackage{graphicx} 
\usepackage{framed}
\usepackage{caption}
\usepackage{subcaption}

\usepackage{booktabs} 
\usepackage{array} 
\usepackage{paralist} 
\usepackage{verbatim} 

\usepackage{lipsum}
\usepackage{setspace}

\usepackage{lipsum}
\usepackage[singlelinecheck=false]{caption}
\usepackage{tabularx}

\newtheorem{assump*}{Assumption}[section]

\theoremstyle{definition}

\def\equationautorefname~#1\null{(#1)\null} 

\let\emptyset\varnothing

\title{Exposure effects are not automatically useful for policymaking} 
\author{\large Eric Auerbach\footnote{Department of Economics, Northwestern University, 2211 Campus Drive, Evanston IL, 60208, eric.auerbach@northwestern.edu} \and Jonathan Auerbach\footnote{Department of Statistics, George Mason University, 4400 University Drive, MS 4A7, Fairfax, VA 22030, jauerba@gmu.edu} \and Max Tabord-Meehan\footnote{Department of Economics, University of Chicago 1126 East 59th Street, Chicago, IL, 60637, maxtm@uchicago.edu}}
\begin{document}
\date{}
\maketitle

\section{Introduction}
We thank \cite{Savje:2023} for a thought-provoking article and appreciate the opportunity to share our perspective as social scientists. In his article, S\"avje recommends misspecified exposure effects as a way to avoid strong assumptions about interference when analyzing the results of an experiment. In this invited discussion, we highlight a limitation of Savje’s recommendation: exposure effects are not generally useful for evaluating social policies without the strong assumptions that Savje seeks to avoid.

Our discussion is organized as follows. Section 2 summarizes our position, Section 3 provides a concrete example, and Section 4 concludes. Proof of claims are in an appendix. 

\section{Exposure effects are not automatically useful}
Exposure effects are commonly used as an organizational tool. The idea is to decompose a complicated interference pattern into individual units that are easier to manage separately. For instance, consider an experimenter who wants to learn whether assigning a treatment to every unit is expected to increase outcomes on average. The effect of this policy may be difficult to estimate directly when there is interference. However, it can often be written as a combination of exposure effects, each of which may be straightforward to estimate using standard techniques. See for instance \cite{Manski:2013} and \cite{Aronow:2017}.

But an individual exposure effect is not automatically relevant or interpretable in isolation. For example, finding that one exposure effect is large and positive does not mean that the treatment is expected to increase outcomes for any collection of units. There may be other negative exposure effects that together effectively cancel out the positive one. Without  accounting for all of the interdependencies between units, an experimenter can never be sure that they have considered every interaction that is relevant for the policy question at hand. It is this accounting exercise that requires  strong assumptions about the interference structure. 

This is why we disagree with S\"avje's recommendation. We do not dispute his technical contribution, which is to show that it is possible to consistently estimate a large class of exposure effects under relatively weak assumptions. However, S\"avje goes on to argue that, as a consequence, experimenters can safely misspecify the interference structure when analyzing the results of an experiment. Our disagreement is that we do not see how the results of an experiment can generally be useful for evaluating social policies outside the context of a correctly specified interference structure.

\section{A concrete example: the effect of additional policing on crime }
To make our position concrete, we consider a hypothetical experiment in which a police department randomly assigns additional officers (the treatment) to city blocks (the units). The goal of the department is to determine the expected number of crimes prevented by the marginal officer. Interference occurs because although the officer may only interact with the people that live and work on that block, those people will travel to other blocks where they may commit or fall victim to a crime. It is important for the department to account for this behavior because it may be, for instance, that the additional officer does not actually prevent any crimes. Rather criminals simply move their operations from the treated blocks to other less-policed areas. If these interactions are ignored, the department may incorrectly conclude that the misplaced crimes were prevented and overstate the expected impact of additional policing.  

In our example, we show that the experimenter can learn about the expected amount of crime prevented by additional policing using a specific combination of exposure effects determined by the interference structure. But without any assumptions, an arbitrary exposure effect can take essentially any sign or magnitude. As a result, it can not possibly be useful for evaluating the impact of additional policing on crime.  

\subsection{Model and policy question}
The amount of crime reported in block $i = 1, \ldots,n$ is given by the model
\begin{align}\label{model}
y_{i}(\boldsymbol{z}) = \alpha_{i} + \beta_{i}z_{i} + \sum_{r=1}^{R}\gamma_{ir}\sum_{j \in N_{i}^{r}}z_{j}
\end{align}
where $z_{i} \in \{0,1\}$ indicates whether an additional officer (treatment) is assigned to block $i$, $\boldsymbol{z} = (z_1, \ldots, z_n) \in \mathcal{Z} := \{0, 1\}^n$, $\alpha_{i}$ is the amount of crime reported in block $i$ when no blocks are treated, $\beta_{i}$ is the change in crime caused by assigning treatment to block $i$, $N_{i}^{r} \subseteq \{1, \ldots, n\}$ is the collection of blocks at distance $r$ from block $i$, $\sum_{j \in N_{i}^{r}}z_{j}$ is the number of treated blocks at distance $r$ from block $i$, and $\gamma_{ir}$ is the change in crime caused by assigning treatment to an additional block of distance $r$ from $i$. In words, $\beta_i$ is a direct effect and $\gamma_{ir}$ is a spillover effect of treatment. The interference structure is determined by the sets $N_{i}^{r}$. We assume $i \not\in N_{i}^{r}$ and $N_{i}^{r}\cap N_{i}^{r'} = \emptyset$ for any $r,r' = 1,...,R$ with $r \neq r'$. Following S\"avje, $y_{i}(\boldsymbol{z})$ is nonstochastic. 

The police department wants to know whether an additional officer is expected to reduce crime. Specifically, they are interested in a treatment policy that assigns the treatment to one block drawn uniformly at random from the city. Under (\ref{model}), the expected amount of crime without the policy is $ \sum_{i}\alpha_{i}$. The expected amount of crime with the policy is $\sum_{i}\alpha_{i} + \bar{\beta} + \bar{\gamma}$ where $\bar{\beta} := \frac{1}{n}\sum_{i}\beta_{i}$ and  $\bar{\gamma} := \frac{1}{n}\sum_{i=1}^{n}\sum_{r = 1}^{R}\gamma_{ir}|N_{i}^{r}|$. The policy is expected to reduce crime if $\bar{\beta} + \bar{\gamma} < 0$. In words, $\bar{\beta} + \bar{\gamma}$ is the average policy effect, $\bar{\beta}$ is the average direct effect, and $\bar{\gamma}$ is the average spillover effect. 

\subsection{The policy question can be addressed using correctly specified exposure effects}
To estimate the average policy effect we express $\bar{\beta}$ and $\bar{\gamma}$ as sums of exposure effects. We show in the supplement that $d_i(\boldsymbol{z}): \mathcal{Z} \rightarrow \Delta$ is a correctly specified exposure mapping where $d_i(\boldsymbol{z}) = (z_{i},\sum_{j \in N_{i}^{1}}z_{j},...,\sum_{j \in N_{i}^{R}}z_{j})$ and $\Delta := \mathbb{R}^{R+1}$. We also show that
\begin{align}
\bar{\beta} &= \frac{1}{n}\sum_{i=1}^{n}\left(\tilde{y}_i(e_1) - \tilde{y}_i(e_0)\right) \label{ef1}\text{ and } \\
\bar{\gamma} &= \sum_{r=1}^{R}\frac{1}{n}\sum_{i=1}^{n}\left(\tilde{y}_i(e_{r+1}) - \tilde{y}_i(e_0)\right)|N_{i}^{r}|\label{ef2}
\end{align}
where $\tilde{y}_i:\Delta \rightarrow \mathbb{R}$ is such that $\tilde{y}_i(d_i(\boldsymbol{z})) = y_i(\boldsymbol{z})$,  $e_{r} \in \mathbb{R}^{R+1}$ has a $1$ in the $r$th entry and a $0$ in every other entry, and $e_{0} \in \mathbb{R}^{R+1}$ is a vector of all $0$s. The exposure effects on the right-hand side of (\ref{ef1}) and (\ref{ef2}) can be directly estimated using data from a randomized experiment. For example, suppose treatment is assigned to block $i$ with probability $p \in (0,1)$ independently across blocks. The random variable $Z_{i} \in \{0, 1\}$ indicates whether the treatment is assigned to block $i$ in the experiment and $\boldsymbol{Z} := (Z_{1},...,Z_{n})$. The Horvitz-Thompson estimators of $\bar{\beta}$ and $\bar{\gamma}$ are 
\begin{align}
\hat{\beta} &=  \frac{1}{n}\sum_{i=1}^n\frac{Y_i\mathbbm{1}\{d_i(\boldsymbol{Z}) = e_1\}}{\pi_i(e_1)} - \frac{1}{n}\sum_{i=1}^n\frac{Y_i\mathbbm{1}\{d_i(\boldsymbol{Z}) = e_0\}}{\pi_i(e_0)}  \\
\hat{\gamma} &= \sum_{r = 1}^R\left(\frac{1}{n}\sum_{i=1}^n\frac{Y_i\mathbbm{1}\{d_i(\boldsymbol{Z}) = e_{r+1}\}}{\pi_i(e_{r+1})} - \frac{1}{n}\sum_{i=1}^n\frac{Y_i\mathbbm{1}\{d_i(\boldsymbol{Z}) = e_0\}}{\pi_i(e_0)}\right)|N_i^r|\label{ef5}
\end{align}
where $Y_i = y_i(\boldsymbol{Z})$ and $\pi_i(d) = \text{pr}(d_i(\boldsymbol{Z}) = d)$. See \cite{Aronow:2017}. This gives an estimator for the average policy effect $\hat{\beta} + \hat{\gamma}$ which can be used to make inferences about the sign of $\bar{\beta} + \bar{\gamma}$ and evaluate whether an additional officer is expected to reduce crime.

\subsection{S\"avje recommends misspecified exposure effects}
The policy evaluation strategy outlined above requires knowledge of the sets $N_{i}^{r}$. In practice, however, the experimenter may not know this interference structure. S\"avje argues that experimenters should in such cases shift their focus to exposure effects that they deem relevant and interpretable, even if they are induced by misspecified exposure mappings. 

To understand S\"avje's recommendation in the context of our example, we suppose the experimenter misspecifies  $N_{i}^{r}$. That is, they assume the interference is given by $\breve{N}_{i}^{r} \subseteq \{1, 2, \ldots, n\}$ that may be different from $N_{i}^{r}$, but still satisfy $i \not \in \breve{N}_{i}^{r} $ and $\breve{N}_{i}^{r}  \cap \breve{N}_{i}^{r'} = \emptyset$ for any $r,r' = 1,...,R$ with $r \neq r'$. Following S\"avje, this misspecification induces the misspecified exposure mappings $\breve{d}_{i}(z) = (z_{i},\sum_{j \in \breve{N}_{i}^{1}}z_{j},...,\sum_{j \in \breve{N}_{i}^{R}}z_{j}) \in \Delta$ and expected potential outcome functions $\bar{y}_i(d): \Delta \rightarrow \mathbb{R}$ where $\bar{y}_i(d) =  E[y_{i}(\boldsymbol{Z})|  \breve{d}_{i}(\boldsymbol{Z}) = d]$. A misspecified average spillover effect is
\begin{align}
\breve{\gamma} &:=  \sum_{r=1}^{R}\frac{1}{n}\sum_{i=1}^{n}\left(\bar{y}_i(e_{r+1}) - \bar{y}_i(e_0)\right)|\breve{N}_{i}^{r}|. \label{ef3}  
\end{align}

\subsection{The policy question can not be addressed using misspecified exposure effects}
Our issue with S\"avje's recommendation is that we do not see how one can meaningfully interpret an exposure effect outside the context of a correctly specified interference structure. In our example, without assumptions about how $\breve{N}_{i}^{r}$ is related to $N_{i}^{r}$, the misspecified average spillover effect $\breve{\gamma}$ can essentially take any sign or magnitude. Specifically, we show in the appendix that $\breve{\gamma}$ is a  weighted combination of spillover effects
\begin{align}
\breve{\gamma} = \frac{1}{n}\sum_{i=1}^{n}\sum_{r=1}^{R}\gamma_{ir}|N_{i}^{r}|\breve{w}_{ir} \text{ where } \breve{w}_{ir} = \sum_{s=1}^{R}\frac{|N_{i}^{r} \cap \breve{N}_{i}^{s}||\breve{N}_{i}^{r}|}{|\breve{N}_{i}^{s}| |N_{i}^{r}|}.
\end{align}
The weights are nonnegative but do not necessarily sum to one. Their magnitude can be arbitrarily large or small, depending on the discrepancy between $N_{i}^{r}$ and $\breve{N}_{i}^{r}$. They can also be arbitrarily related to the sign of $\gamma_{ir}$, so that the sign and magnitude of $\breve{\gamma}$ may be completely unrelated to the sign or magnitude of the average policy, direct, or spillover effect. 

We can think of two specific settings where $\breve{\gamma}$, or any other misspecified exposure effect, would be useful. The first setting is when the researcher is narrowly interested in testing the hypothesis that all of the spillover effects $\gamma_{ir}$ are zero. Estimating a nonzero $\breve{\gamma}$ provides evidence against this hypothesis. The second setting is the case where the spillover effects all have the same sign. Under this strong assumption, the sign of $\breve{\gamma}$ reveals the sign of the spillovers.

However, we do not see how misspecified exposure effects can generally be useful for policy making. They can be small in magnitude when the expected impact of the policy is large. They can also be large when the impact of the policy on every collection of units is small. Only with knowledge of the interference structure can the experimenter take exposure effects and, using formulas akin to (\ref{ef1}) and (\ref{ef2}), say something concrete about the policy of interest.

\section{Conclusion}
Our main purpose in writing this discussion is to point out that exposure effects are not automatically useful without assumptions on the interference structure. However, misspecified exposure effects may still serve as useful approximations for some policy effects of interest. For instance, in our own work, \cite{Auerbach:2021} develop an exposure mapping ``sieve" based on rooted networks. We show that a large class of network interference patterns can be  well approximated by a specific sequence of exposure mappings. These exposure mappings are misspecified in finite samples, but because they are correctly specified asymptotically, we are able to consistently estimate various policy effects. However, the quality of our approximation depends crucially on how closely our exposure map sieve approximates the true interference structure and this plays a role in determining the statistical properties of our proposed estimators.  

Ultimately, there is no free lunch. Causal inference with interference is a complex problem. Experimenters must take a stance on the structure of the interference in their experiments if they want to characterize the impact of a treatment in a way that is useful for policy making. 

\section*{Acknowledgement}
Research is supported by NSF grants SES-2149408 and SES-2149422.

\appendix

\section*{Proof of claims}
Recall that 
\begin{align*}
y_{i}(\boldsymbol{z}) &= \alpha_{i} + \beta_{i}z_{i} + \sum_{r=1}^{R}\gamma_{ir}\sum_{j \in N_{i}^{r}}z_{j}, \\
\bar{\beta} &:= \frac{1}{n}\sum_{i}\beta_{i}, \text{ and } \\
\bar{\gamma}  &:= \frac{1}{n}\sum_{i=1}^{n}\sum_{r = 1}^{R}\gamma_{ir}|N_{i}^{r}|. 
\end{align*}
Our first claim is 
\begin{flushleft}
\textbf{Claim 1:} $d_{i}(\boldsymbol{z}) = (z_{i},\sum_{j \in N_{i}^{1}}z_{j},...,\sum_{j \in N_{i}^{R}}z_{j}) \in \mathbb{R}^{R+1}$ is an exposure map. 
\end{flushleft}
\begin{flushleft}
\textbf{Proof of claim 1:} $y_{i}(\boldsymbol{z})$ is a deterministic function of $d_{i}(\boldsymbol{z})$ so that for any $\boldsymbol{z}, \boldsymbol{z}'  \in \{0,1\}^{n}$,  $d_{i}(\boldsymbol{z}) = d_{i}(\boldsymbol{z}')$ implies that $y_{i}(\boldsymbol{z}) = y_{i}(\boldsymbol{z}')$. $\square$
\end{flushleft}
Our second claim is
\begin{flushleft}
\textbf{Claim 2:} $\bar{\beta} = \frac{1}{n}\sum_{i=1}^{n}\left(\tilde{y}_i(e_1) - \tilde{y}_i(e_0)\right)$ and 
$\bar{\gamma} = \sum_{r=1}^{R}\frac{1}{n}\sum_{i=1}^{n}\left(\tilde{y}_i(e_{r+1}) - \tilde{y}_i(e_0)\right)|N_{i}^{r}|.$
\end{flushleft}
\begin{flushleft}
\textbf{Proof of claim 2:} $d_{i}(\boldsymbol{z}) = e_{0}$ implies $z_{i} = 0$ and $\sum_{j\in N_{i}^{s}}z_{j} = 0$ for $s = 1,...,R$, so that $\tilde{y}_i(e_0) = \alpha_{i}$. $d_{i}(\boldsymbol{z}) = e_{1}$ implies $z_{i} = 1$ and $\sum_{j\in N_{i}^{s}}z_{j} = 0$ for $s = 1,...,R$ so that $\tilde{y}_i(e_1) = \alpha_{i} + \beta_{i}$. $d_{i}(\boldsymbol{z}) = e_{r+1}$ for $r = 1,...,R$ implies $z_{i} = 0$, $\sum_{j\in N_{i}^{s}}z_{j} = 1$ for $s = r$, and $\sum_{j\in N_{i}^{s}}z_{j} = 0$ for $s \not= r$ so that $\tilde{y}_i(e_{r+1}) = \alpha_{i} + \gamma_{ir}$. $\square$
\end{flushleft}
Our third claim is 
\begin{flushleft}
\textbf{Claim 3:} $\breve{\gamma} = \frac{1}{n}\sum_{i=1}^{n}\sum_{r=1}^{R}\gamma_{ir}|N_{i}^{r}|\breve{w}_{ir}$ where $\breve{w}_{ir} = \sum_{s=1}^{R}\frac{|N_{i}^{r} \cap \breve{N}_{i}^{s}||\breve{N}_{i}^{r}|}{|\breve{N}_{i}^{s}| |N_{i}^{r}|}$.
\end{flushleft}
\begin{flushleft}
\textbf{Proof of claim 3:} Define  $\breve{N}_{i} := \cup_{r =1}^{R}\breve{N}_{i}^{r}$ and write $y_{i}(\boldsymbol{Z})$ as
\begin{align*}
y_{i}(\boldsymbol{Z}) = \alpha_{i} + \beta_iZ_i + \sum_{r=1}^{R}\gamma_{ir}\sum_{j \in N_{i}^{r}}Z_{j}\mathbbm{1}\{j \not\in \breve{N}_{i}\} + \sum_{r=1}^{R}\gamma_{ir}\sum_{j \in N_{i}^{r}}Z_{j}\mathbbm{1}\{j \in \breve{N}_{i}\}.
\end{align*}
The event $\breve{d}_{i}(\boldsymbol{Z}) = e_{0}$ implies that $Z_{j} = 0$ for every $j \in \breve{N}_{i} \cup \{i\}$ and so $ \sum_{r=1}^{R}\gamma_{ir}\sum_{j \in N_{i}^{r}}Z_{j}\mathbbm{1}\{j \in \breve{N}_{i}\} = 0$. It follows that 
\begin{align*}
E[y_{i}(\boldsymbol{Z})| \breve{d}_{i}(\boldsymbol{Z}) = e_{0}] = \alpha_{i} +  \sum_{r=1}^{R}\gamma_{ir}\sum_{j \in N_{i}^{r}}\mathbbm{1}\{j \not\in \breve{N}_{i}\}p
\end{align*}
 because the entries of $\boldsymbol{Z}$ are iid Bernoulli($p$). It follows from the same logic that 
 \begin{align*}
E[y_{i}(\boldsymbol{Z})| \breve{d}_{i}(\boldsymbol{Z}) = e_{s}] = \alpha_{i} +  \sum_{r=1}^{R}\gamma_{ir}\sum_{j \in N_{i}^{r}}\mathbbm{1}\{j \not\in \breve{N}_{i}\}p  + \sum_{r=1}^{R}\gamma_{ir}\breve{w}_{irs}
\end{align*}
for $s = 2,...,R+1$ where $\breve{w}_{irs} := E\left[\sum_{j \in N_{i}^{r}}Z_{j}\mathbbm{1}\{j \in \breve{N}_{i}\} |\breve{d}_{i}(\boldsymbol{Z}) = e_{s}\right]$. As a result, $\breve{\gamma} = \frac{1}{n}\sum_{i=1}^{n}\sum_{r=1}^{R}\gamma_{ir}|N_{i}^{r}|\breve{\omega}_{ir}$ where $\breve{\omega}_{ir} := \sum_{s=1}^{R}\breve{w}_{irs}|\breve{N}_{i}^{r}|/ |N_{i}^{r}|$. What remains to be shown is that $\breve{w}_{irs} = |N_{i}^{r} \cap \breve{N}_{i}^{s}|/|\breve{N}_{i}^{s}|$. To demonstrate this, we note that that under the event $\breve{d}_{i}(\boldsymbol{Z}) = e_{s}$ for $s = 2,...,R+1$, exactly one unit in $\breve{N}_{i}^{s}$ is treated and no units in $\breve{N}_{i}^{s'}$ are treated for $s' \neq s$. $\breve{w}_{irs} = E\left[\sum_{j \in N_{i}^{r}}Z_{j}\mathbbm{1}\{j \in \breve{N}_{i}\} |\breve{d}_{i}(\boldsymbol{Z}) = e_{s}\right]$ is then just the probability that the one treated unit in $\breve{N}_{i}^{s}$ is also an element of $N_{i}^{r}$. Since the entries of $\boldsymbol{Z}$ are iid, each element in $\breve{N}_{i}^{s}$ is equally likely to be the treated one, and so the probability that the treated unit is in $N_{i}^{r}$ is proportional to $|N_{i}^{r} \cap \breve{N}_{i}^{s}|$. It follows that $\breve{w}_{irs} = |N_{i}^{r} \cap \breve{N}_{i}^{s}|/|\breve{N}_{i}^{s}|$. $\square$
\end{flushleft}

\end{document}